\documentclass{pasj01}
\Received{$\langle$reception date$\rangle$}
\Accepted{$\langle$acception date$\rangle$}
\Published{$\langle$publication date$\rangle$}

\begin{document}

\title{Gravitational waves from an SMBH binary in M87}
\author{
Naoyuki \textsc{Yonemaru}\altaffilmark{1},
Hiroki \textsc{Kumamoto}\altaffilmark{1},
Sachiko \textsc{Kuroyanagi}\altaffilmark{2},
Keitaro \textsc{Takahashi}\altaffilmark{1},
Joseph \textsc{Silk}\altaffilmark{3,4}
}%
\altaffiltext{1}{Kumamoto University, Graduate School of Science and Technology, Japan}
\altaffiltext{2}{Nagoya University, Graduate School of Science, Japan}
\altaffiltext{3}{Institut d'Astrophysique de Paris, UMR 7095, CNRS, UPMC Universit\'{e} Paris 6, Sorbonne Universit\'{e}s, 98 bis boulevard Arago, 75014 Paris, France and  AIM-Paris-Saclay, CEA/DSM/IRFU, CNRS, Univ. Paris VII, F-91191 Gif-sur-Yvette, France}
\altaffiltext{4}{The Johns Hopkins University, Department of Physics and Astronomy, 3400 N. Charles Street, Baltimore, Maryland 21218, USA}

\email{157d8019@st.kumamoto-u.ac.jp}


\maketitle

\begin{abstract}
In this paper, we study gravitational-wave (GW) emission from a hypothetical supermassive black-hole (SMBH) binary at the center of M87. The existence of a SMBH other than that usually identified with the central  AGN is a possible explanation for the observed displacement ($\sim O(1)~{\rm pc}$) between the AGN and the galactic centroid, and it is reasonable to assume considering the evolution of SMBHs through galaxy mergers. Because the period of the binary and the resulting GWs is much longer than the observational time span, we calculate the variation of the GW amplitude, rather than the amplitude itself. We investigate the dependence on the orbital elements and the second BH mass taking the observational constraints into account. The frequency of the GWs is too low to be detected with the conventional pulsar timing array and we propose a new method to detect such low-frequency GWs with the distribution function of pulsar spin-down rates. Although the GWs from a SMBH binary which explains the observed displacement is extremely hard to be detected even with the new method, GWs are still a useful way to probe the M87 center.
\end{abstract}

\section{Introduction}
 
A giant elliptical galaxy M87 (NGC 4486), located at 18.4 Mpc, hosts one of the nearest and best-studied AGN. A relativistic one-sided jet is ejected from the galactic center, where a supermassive black hole (SMBH) with a mass of $(6.6 \pm 0.4) \times 10^9~M_\odot$ \citep{Gebhardt} resides, and extends to hundreds of kilo parsecs.

It was reported in \citet{Batcheldor} that the luminous  center of M87 and the AGN are displaced significantly. The displacement was confirmed in \citet{Lena2014} and this fact suggests that the SMBH is not located at the center of mass of the galaxy, although there is a large uncertainty in the projected distance ($\sim 1 - 7~{\rm pc}$). Several scenarios have been suggested to interpret this displacement: (1) accelaration by an asymmetric jet, (2) recoil due to gravitational-wave emission caused by a merger of two SMBHs, (3) binary motion with respect to another SMBH and (4) gravitational perturbations from multiple massive objects at the galactic center. As a possible method to distinguish between these scenarios,  observation of the proper motion of the SMBH (AGN) was suggested.

Here we focus on the latter two scenarios. Binary formation of two SMBHs through galaxy mergers is a basic process of SMBH evolution. The observed displacement can be explained if the center of mass of a SMBH binary is located at the galactic center and only one of the SMBHs has enough gas accretion to make it an AGN. On the other hand, there is a possibility that multiple intermediate mass BHs (IMBHs) or relatively small SMBHs reside at the galactic center through hierarchical galaxy merging \citep{Islam2004,Rashkov2014}. The displacement can also be explained by gravitational perturbation of the main SMBH from these multiple BHs. In this latter scenario, the main SMBH can form a binary with one of the IMBHs or SMBHs, and the center of mass will not be necessarily located at the galactic center.

If the observed displacement is caused by these mechanisms, a substantial amount of gravitational waves (GWs) is emitted by the SMBH binary. Thus, observation of GWs from M87 can probe the black hole population at the galactic center. In this paper, following these scenarios, we estimate the GWs under the constraints on the binary orbital elements from observations and discuss the detectability of GWs of pulsar timing arrays.

This paper is organized as follows. In section 2, we summarize the basic formulae of Keplerian motion. In section 3, we describe the formalism to calculate the amplitude of GWs from a binary and its time derivative. In section 4, we present our main results. In section 5, we discuss the detectability of the GWs and give a summary and discussion in section 6. For the rest of this paper we set $c = G = 1$, unless otherwise specified.

\section{Orbital elements of a SMBH binary}

The distance between the main SMBH and the barycenter of its host galaxy, $r_1$, is related to the projected displacement, $r_0$, as,
\begin{equation}
r_1 = \frac{1}{\sin{\theta}}\ r_0\ ,
\end{equation}
where $\theta$ is the angle between the line of sight and the direction from the SMBH to the barycenter.

In the following, we discuss the orbital elements of the SMBH binary. The orbital motion is approximated to the Newtonian two-body problem since the binary has a typical separation of $\gtrsim 1~{\rm pc}$ so that gravity is sufficiently weak and the energy loss due to gravitational waves is negligible. The equation of motion for the relative motion of the SMBHs is given by
\begin{equation}
\mu\,\frac{d^2\vec{r}}{dt^2} = -\frac{G m_1 m_2}{r^2}\frac{\vec{r}}{r}\ ,
\end{equation}
where $\vec{r} = \vec{r}_1 - \vec{r}_2$, $r = |\vec{r}|$, $\mu = m_1 m_2/(m_1 + m_2)$ is the reduced mass, and $m_1 = 6.6 \times 10^9 ~M_\odot$ and $m_2$ are the mass of the main and second SMBH, respectively. The orbital elements are described by
\begin{eqnarray}
r &=& \frac{p}{1 + e\,\cos{\phi}} \label{eq:r} \\
r^2 \frac{d\phi}{dt} &=& \bigl( M p \bigr)^{1/2}\ , \label{eq:dphidt}
\end{eqnarray}
where $p = a( 1 - e^2 )$, $e$ is the eccentricity ($0 \leq e<1$), $a$ is the semi-major axis, $\phi$ is the orbital phase and $M = m_1 + m_2$. Fixing the eccentricity, the semi-major axis is maximum (minimum) when the SMBH is located at the pericenter (apocenter) of the orbit. Therefore, the semi-major axis is constrained as  
\begin{equation}
\frac{r}{1+e}\ \leq a \leq\ \frac{r}{1-e}\ .
\end{equation}
On the other hand, the orbital period is written by 
\begin{equation}
T = 2 \pi \sqrt{\frac{a^3}{M}}\ .
\end{equation}
In the current case, the typical orbital period is $2 \times 10^3$ yrs for $a_1 = 1~{\rm pc}$ and $m_2 = 6.6 \times 10^9 M_{\odot} (= m_1)$.

\section{Gravitational waves from a binary}

Gravitational waves are emitted from a binary and the waveforms of the plus and cross polarizations are given by,
\begin{eqnarray}
h_+(t) &=& \frac{1}{2}\,\biggl(p^ip^j - q^iq^j\biggr)\ h^{TT}_{ij}
\label{eq:hplus} \\
h_{\times}(t) &=& \frac{1}{2}\,\biggl(p^iq^j + p^jq^i\biggr)\ h^{TT}_{ij}\ .
\label{eq:hcross}
\end{eqnarray}
Here, $p = (0,1,0)$ and $q = (-\cos\iota,0,\sin\iota)$ are the unit vectors of the gravitational wave principal axes where $\iota$ is the inclination and $h^{TT}_{ij} (i,j = 1,2,3)$ is the transverse-traceless part of the amplitude given by \citep{Gopakumar}
\begin{equation}
h^{TT}_{ij} = \frac{4\mu}{R}\ \biggl(v_iv_j -  \frac{M}{r}n_in_j\biggr)\ ,
\end{equation}
where $n = (\rm{cos}\phi, sin\phi, 0)$ and $v = (\dot{r}{\rm{cos}}\phi - r\dot{\phi}{\rm{sin}}\phi, \dot{r}{\rm{sin}}\phi + r\dot{\phi}{\rm{cos}}\phi, 0)$. Eqs. (\ref{eq:hplus}) and (\ref{eq:hcross}) can be rewritten as,
\begin{eqnarray}
h_+(t) &=& \frac{\mu}{R}\biggl\{ (1+{\rm{cos}}^2\iota ) \biggl[ \biggl(\frac{M}{r} + r^2\dot{\phi}^2 - \dot{r}^2\biggr) {\rm{cos}}2\phi \nonumber \\
&& + 2r\dot{r}\dot{\phi}\ {\rm{sin}}2\phi \biggr] + \biggl[\frac{M}{r} - r^2\dot{\phi}^2 -\dot{r}^2\biggr]{\rm{sin}}^2\iota\biggr\} \\
h_{\times}(t) &=& \frac{2\mu}{R}{\rm{cos}}\,\iota \nonumber \\
&& \times
   \left\{ \left( \frac{M}{r}+r^2\dot{\phi}^2 - \dot{r}^2 \right)
           \sin{2\phi} - 2 r \dot{r} \dot{\phi} \cos{2 \phi}
   \right\} ,
\end{eqnarray}
where $R = 18.4~{\rm Mpc}$ is the distance to M87.

Assuming the photo-center of the galaxy is on the orbital plane, the inclination is constrained as
\begin{eqnarray}
90^\circ -\theta\leq\iota\leq90^\circ + \theta, \\
\theta-90^\circ \leq\iota\leq270^\circ - \theta
\end{eqnarray}
in the case where the main SMBH is located in front of ($\theta\leq 90^\circ$) and behind ($\theta> 90^\circ$) the photo-center, respectively.

In the current situation, the period of gravitational waves is much longer than the observation time span of pulsar timing arrays ($O(10)~{\rm years}$) so that the amplitude varies linearly with time on this time-scale. Thus, we consider the change in the amplitude in the observational time span, rather than the amplitude itself. To do this, we need the time derivative of the waveforms:
\begin{eqnarray}
\frac{dh_+(t)}{dt}
&=& \frac{\mu}{R}\ \biggl\{ (1+{\rm{cos}}^2\iota)\biggl[\biggl(\frac{M}{r^2}+2r\dot{\phi}^2\biggr)\ \dot{r}\,{\rm{cos}}2\phi 
\nonumber \\
&& -4\frac{M}{r}\dot{\phi}\ {\rm{sin}}2\phi\biggr] - \frac{M}{r^2}\dot{r}\ {\rm{sin}}^2\iota\biggr\} \label{eq:dhdt_plus} \\
\frac{dh_{\times}(t)}{dt}
&=& \frac{2 \mu}{R} \cos{\iota}
    \left\{ 4 \left(\frac{M}{r} - \dot{r}^2\right) \dot{\phi} \cos{2\phi}
            + \frac{M}{r^2} \dot{r} \sin{2 \phi} \right\} .
\label{eq:dhdt_cross}
\end{eqnarray}   
These time derivatives can be expressed in terms of $r_1$, $e$, $a_1$ and $m_2$ by substituting Eqs. (\ref{eq:r}) and (\ref{eq:dphidt}) to Eqs. (\ref{eq:dhdt_plus}) and (\ref{eq:dhdt_cross}).

\section{Results}

\begin{figure}
\includegraphics[width=60mm, angle=-90]{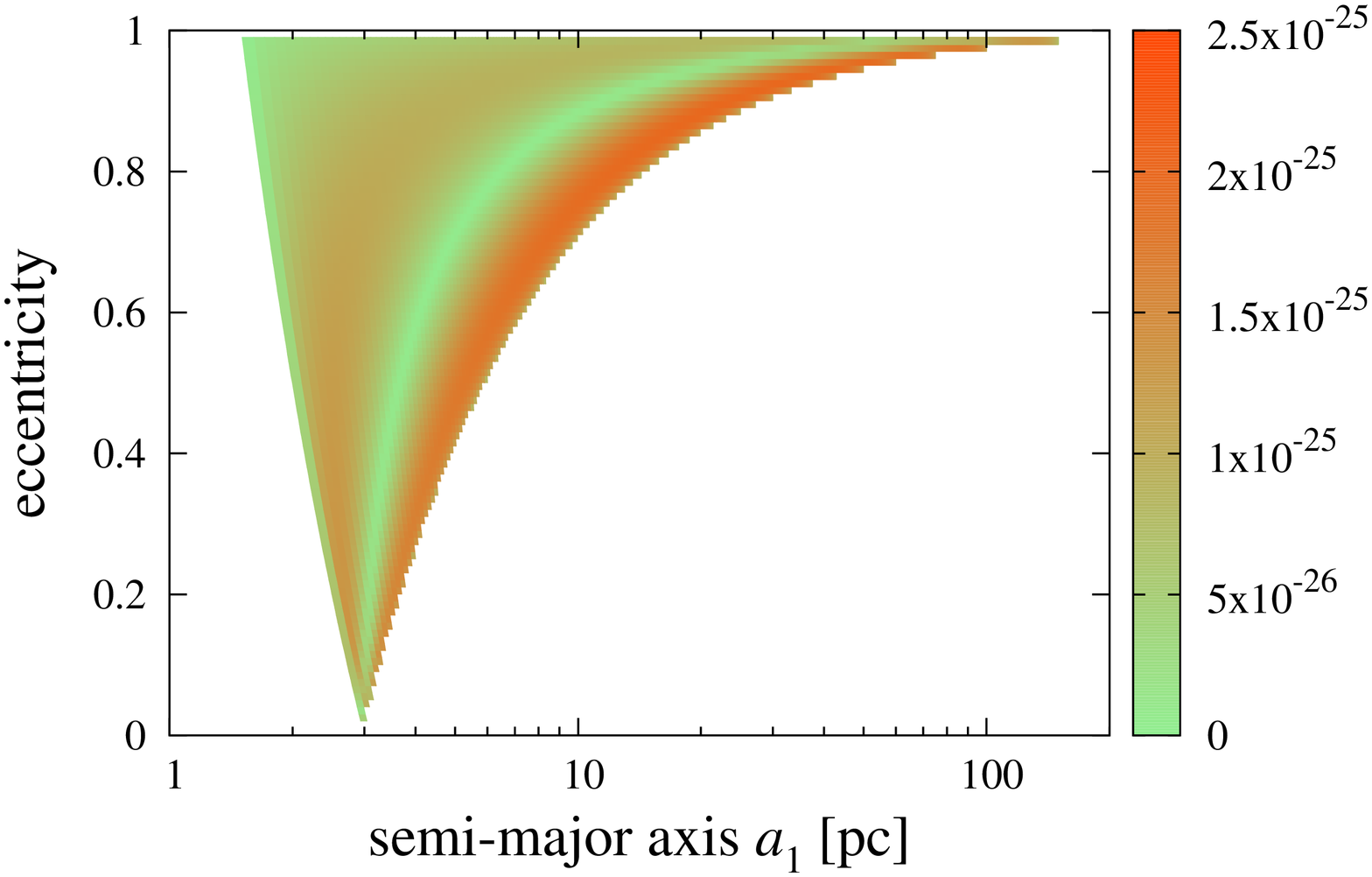}
\includegraphics[width=60mm, angle=-90]{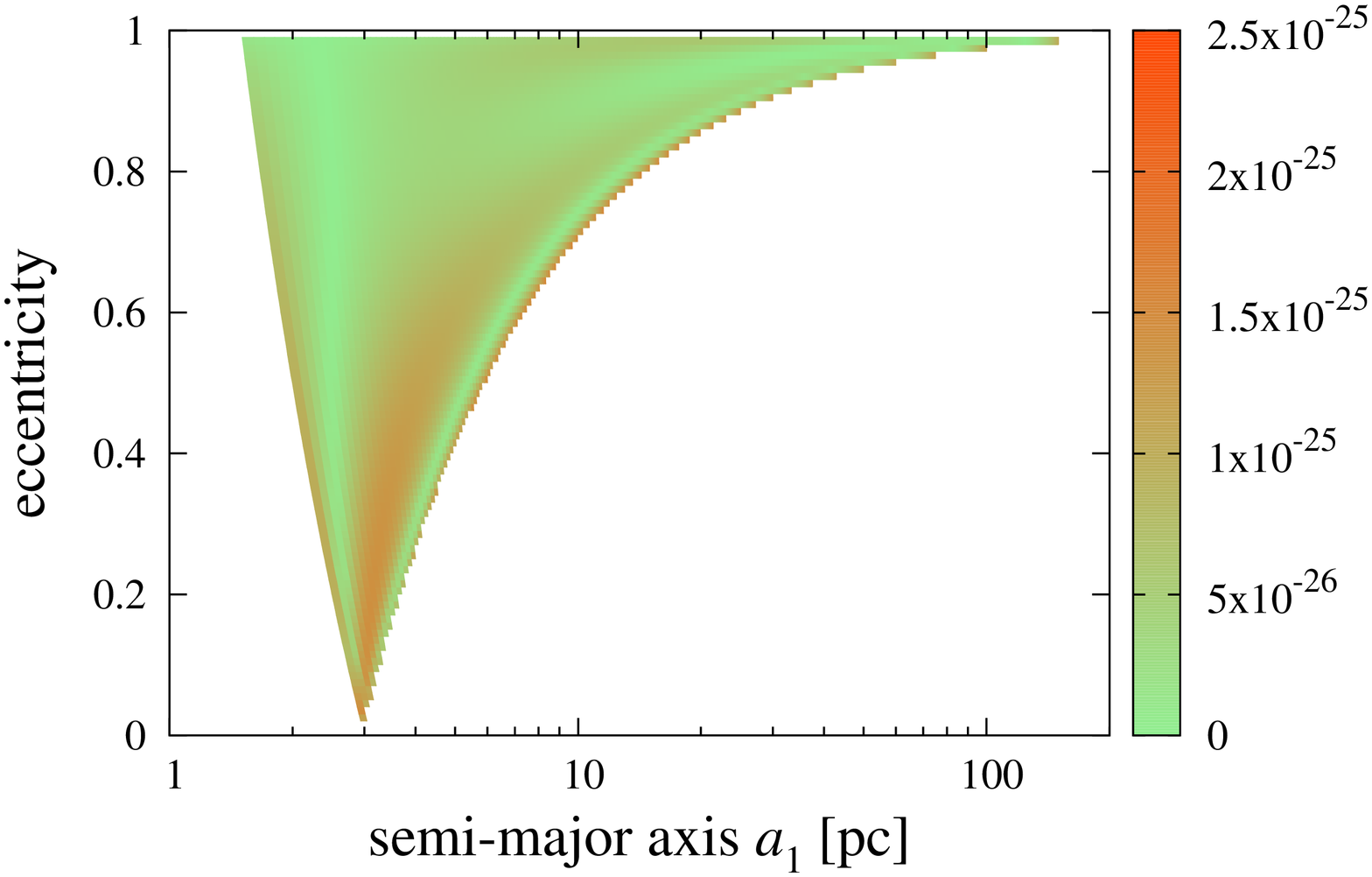}
\vspace{-0.5mm}
\caption{Absolute value of the time derivative of the plus-mode (top) and cross-mode (bottom) amplitudes for $r_1 = 3~{\rm pc}$ and $m_2 = m_1$.}
\label{fig:contour}
\end{figure}

\begin{figure}
\includegraphics[width=55mm, angle=-90]{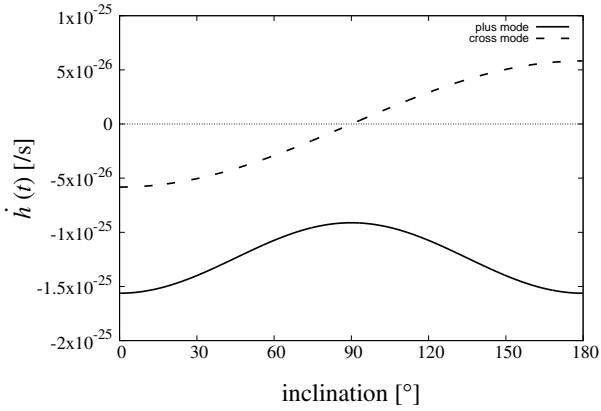}
\vspace{1mm}
\caption{Inclination dependence of the time derivatives (change of the gravitational-wave amplitude in $10~{\rm years}$) for $r_1 = 3~{\rm pc}$, $e = 0.8$, $a_1 = 10.0~{\rm pc}$ and $m_2 = m_1$.}
\label{fig:inclination}
\end{figure}

First, we show the results for the case where the center of mass of two SMBHs is located at the photo-center of M87. Fig. \ref{fig:contour} represents the absolute value of the time derivatives of the plus-mode and cross-mode amplitudes in $a_1-e$ plane for $r_1 = 3~{\rm pc}$ and $m_2 = m_1$. Hereafter, the inclination is set to $0^\circ$ except for Fig. \ref{fig:inclination}. The time derivative is of order $10^{-25}~{\rm sec}^{-1}$ irrespective of the values of $a_1$ and $e$, although the ratio of the two polarization modes depends significantly on the phase of the orbital motion $\phi$, which is uniquely determined by $a_1$ and $e$.

\begin{figure}
\includegraphics[width=55mm, angle=-90]{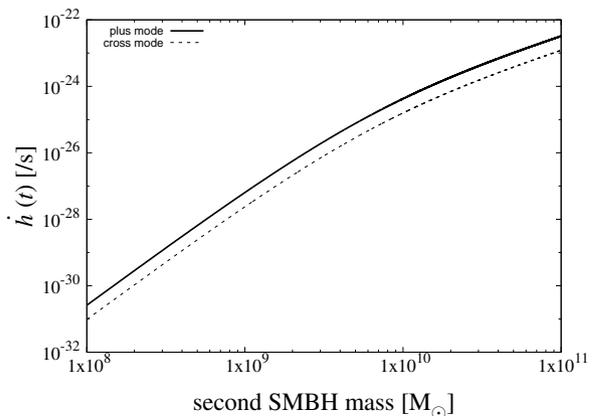}
\vspace{0.5mm}
\caption{Dependence of the time derivatives on the mass of the second SMBH for $r_1 = 3~{\rm pc}$, $e = 0.8$ and $a_1 = 10~{\rm pc}$.}
\label{fig:mass}
\end{figure}

Fig. \ref{fig:mass} shows the dependence of the time derivatives on the mass of the second SMBH for $r_1 = 3~{\rm pc}$, $e = 0.8$ and $a_1 = 10.0~{\rm pc}$. The dependence is strong ($\propto m_2^{3.5}$) for $m_2 < m_1$ and becomes relatively weak ($\propto m_2$) for $m_2 > m_1$.

\begin{figure}
\includegraphics[width=55mm, angle=-90]{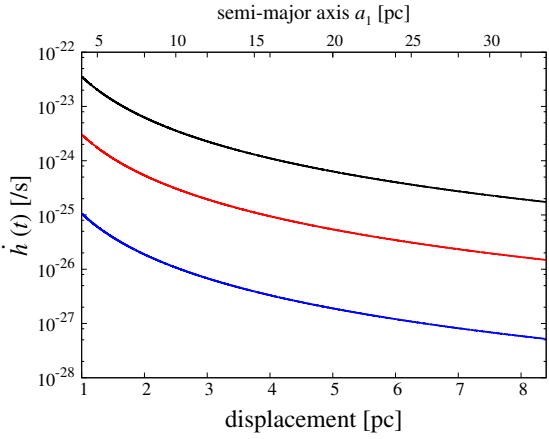}
\vspace{2mm}
\caption{Dependence of the time derivatives on the displacement (projected distance) for $e = 0.8$, $\phi = 56.6^\circ$. The curves are for $m_2 = 3 m_1, m_1$ and $0.3 m_1$ from top to bottom. The upper axis shows the semi-major axis of the main SMBH which corresponds to the displacement of the lower axis fixing eccentricity and the phase of the orbital motion.}
\label{fig:displacement}
\end{figure}

Next, considering that the  observed displacement has a large uncertainty, we show the dependence of the same quantity as Fig. \ref{fig:mass} on the displacement in Fig. \ref{fig:displacement}. The mass of the second SMBH is set to $3 m_1, m_1$ and $0.3 m_1$. From the figure, we see that the change of GW amplitude increases drastically as the displacement decreases. The current uncertainty in the displacement leads to the uncertainty in the change of about 4 orders of magnitude. Here it should be noted that the correspondence between the displacement and the semi-major axis of the main SMBH, which is shown in the upper axis, depends on the eccentricity, which is set to 0.8 in this figure.

Fig. \ref{fig:inclination} shows the inclination dependence fixing all the other parameters. The inclination does not affect the overall amplitude so much, although the derivative of the cross-mode amplitude becomes zero at $90^{\circ}$.

\begin{figure}
\includegraphics[width=55mm, angle=-90]{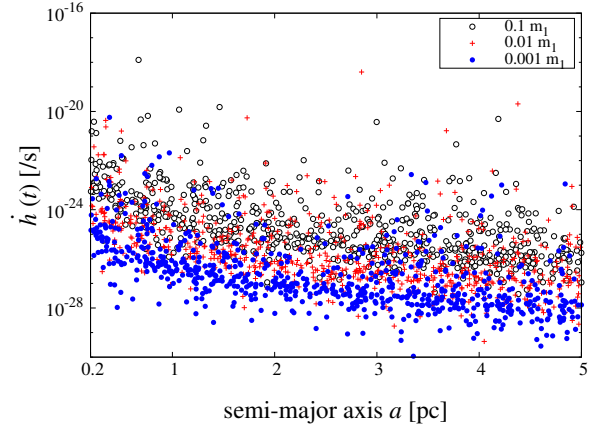}
\vspace{2mm}
\caption{Time derivative of GW amplitude for random sets of orbital elements (semi-major axis, eccentricity, phase and inclination), fixing the mass of the smaller BH. The results are shown as a function of the semi-major axis for $m_2 = 0.1 m_1, 0.01 m_1$ and $0.001 m_1$.}
\label{fig:random}
\end{figure}

Finally, we consider the second scenario where multiple BHs reside in the central region of M87 and the main SMBH is forming a binary with a smaller BH. As remarked in section 1, the center of mass of the binary need not be located at the galactic center and the binary can have a much smaller semi-major axis compared with the previous cases. Because there is no constraint on the orbital elements, we compute the time derivative of GW amplitude for a number of random sets of orbital elements (semi-major axis, eccentricity, phase and inclination), fixing the mass of the smaller BH. The results are shown as a function of the semi-major axis in Fig. \ref{fig:random} for $m_2 = 0.1 m_1, 0.01 m_1$ and $0.001 m_1$. Aside from $m_2$ and semi-major axis, the time derivative of GW amplitude has a strong dependence on the eccentricity and increases significantly for $e \gtrsim 0.8$.

\section{Detectability of Gravitational Waves}

In this section, let us discuss the detectability of the GWs from a possible SMBH binary at the center of M87. Nano-hertz GWs from SMBH binaries are targets of so-called pulsar timing arrays \citep{Sesana2008,Hobbs2010,Lee2011,Mingarelli2013,Sesana2013,Shannon2013,Taylor2013,Burke2015,Janssen2015}. Because the arrival times of pulses are modulated by GWs, the time variation of timing residuals, the difference between the actual arrival time and the expectation in the absence of GWs, follows the wave form of the GWs crossing the earth. Thus, precise measurements of timing residuals can detect GWs with a long period comparable to the time span of observation, typically $O(10)~{\rm years}$.

In the current case with the linearly-changing GWs, the situation is totally different as we see below. The timing residual as a function of time, $r_{\rm GW}(t)$, induced by the GWs is generally described as,
\begin{equation}
r_{\rm GW}(t)
= \frac{1}{2}\frac{\hat{p}^i\hat{p}^j}{1+\hat{\Omega}\cdot\hat{p}}
  \int^{t} \Delta h_{ij}(t',\hat{\Omega}) dt'\label{eq:resh1} .
\end{equation}
where $\hat{\Omega}$ is the direction of the GW propagation, $\hat{p}$ is the unit vector of the direction of the pulsar, and $\Delta h_{ij}(t,\Omega)$ is the difference between the metric perturbation at the earth and pulsar generated by GWs:
\begin{equation}
\Delta h_{ij} = h_{+,\times}(t,\hat{\Omega}) - h_{+,\times}(t_p,\hat{\Omega}) ,
\end{equation}
where $t_p = t - L/c$ and $L$ is the distance to the pulsar. Below, we consider only the first term, so called Earth term, and neglect the second term, pulsar term, following previous works. Assuming that the GWs are linearly changing, that is, $\dot{h}_{+,\times}$ are constant for an observation period, we have,
\begin{equation}
\Delta h_{ij}
= \left(\dot{h}_+ e^+_{ij} + \dot{h}_\times e^\times_{ij} \right)t ,
\end{equation}
where $e^{+,\times}_{ij}$ are the polarization tensors. Then, Eq. (\ref{eq:resh1}) is written as,
\begin{equation}
r_{\rm GW}(t)
= \frac{1}{2} \sum_{A=+,\times} F^A(\hat{\Omega}) \dot{h}_A t^2 \label{eq:resh2} ,
\end{equation}
where $F^A(\hat{\Omega})$ is the antenna beam pattern and given by
\begin{equation}
F^A(\hat{\Omega}) = \left[\frac{1}{2}\frac{\hat{p}^i\hat{p}^j}{1+\hat{\Omega}\cdot\hat{p}}e^A_{ij}(\hat{\Omega})\right] .
\end{equation}
Thus, linearly changing GWs cause timing residuals quadratic in time. However, this is exactly the same behavior as the one due to the spin down of the pulsar,
\begin{equation}
r_{\dot{p}} = \frac{1}{2} \frac{\dot{p}}{p} t^2 \label{eq:resp} .
\end{equation}
where $p$ is the pulse period and $\dot{p}$ is the spin-down rate. Therefore, the linearly changing GWs are absorbed by the spin-down rate in the fitting of timing residual data with a timing model. In other words, the value of $\dot{p}$ is biased by $\alpha(\hat{\Omega}) \equiv \sum_A F^A(\hat{\Omega}) \dot{h}_A$, and the effects of GWs are absorbed as $\alpha (\hat{\Omega})p$.


It should be noted that $\alpha (\hat{\Omega})$ can be positive or negative depending on the position of the pulsar and the GW polarization, while it is natural to consider an intrinsic $\dot{p}$ should be positive. However, some of the observed pulsars have a negative $\dot{p}$ value due to various effects such as acceleration of the pulsar along the line of sight, the differential Galactic rotation and the Shklovskii effect \citep{Shklovskii1970}. Thus, we need to investigate the statistical feature of $\dot{p}$ of many pulsars, rather than that of individual pulsars, in order to probe the existence of linearly-changing GWs.

\begin{figure}[t]
\includegraphics[width=60mm, angle=-90]{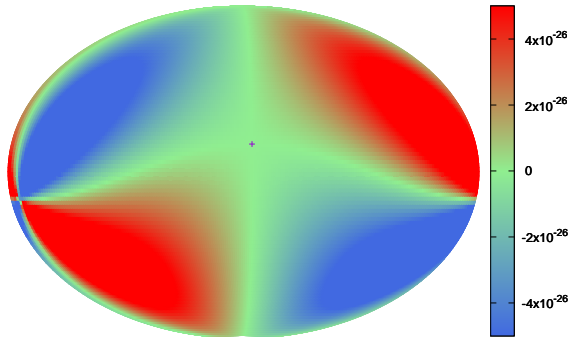}
\caption{Plot of the bias factor $\alpha(\hat{\Omega})$ in the sky for $\dot{h}_+ = 0$ and $\dot{h}_\times = 10^{-25}$, which shows the antenna beam pattern for the cross mode. The symbol "$+$" in the figure represents the position of M87.}
\label{fig:hdot}
\end{figure}

Here, we propose the distribution function of pulsar spin-down rate as a more robust method to detect linearly changing GWs. As noted above, the value of the bias $\alpha(\hat{\Omega})$ depends on the pulsar position and the GW polarization. An example is shown in Fig. \ref{fig:hdot} for $\dot{h}_+ = 0$ and $\dot{h}_\times = 10^{-25}$, which shows the antenna beam pattern in the sky for the cross mode. As can be seen, the positive and negative areas appear as a quadrupole pattern. Thus, if we make histograms of $\dot{p}$ of the pulsars in the positive and negative regions of the sky separately, a systematic difference between the two histograms would be seen for $\dot{p} \lesssim |\alpha(\hat{\Omega})p|$. Practically, because we do not know the polarization of the GWs from M87, we need to compare two histograms assuming any possible polarization and dividing the sky accordingly.

The typical value of $\dot{p}$ of the millisecond pulsars which have been found so far is about $10^{-20}$ and $\dot{p}/p$ can be as low as $10^{-19}$ \citep{Ho2014}. Therefore, linearly changing GWs with $\dot{h} \gtrsim 10^{-19}$ could be, in principle, detected by the above method if we have a sufficiently large sample of millisecond pulsars. According to our calculation above, this level of GWs would not be generated in the first scenario (see Fig. \ref{fig:displacement}), unless the second black-hole mass is extremely large ($\gg 10 m_1$). On the other hand, there is a small chance for the second scenario (see Fig. \ref{fig:random}), if the second BH mass is as large as $0.1 m_1$, the semi-major axis is relatively small $\sim 1~{\rm pc}$ and the eccentricity is large.

In practice, due to the limited number of millisecond pulsars, the statistical errors in the histograms diminish the sensitivity of this method. The number of known millisecond pulsars is approximately 200 (the ATNF pulsar catalogue\footnote{http://www.atnf.csiro.au/people/pulsar/psrcat/}) and will reach 1,400 and 3,000 by the planned SKA1 and SKA2 surveys, respectively \citep{Keane2015,Kramer2015}. It is beyond the scope of the current paper to estimate the sensitivity quantitatively taking the statistical errors and other observational systematics into account.

Let us comment on another aspect of this method which is significantly different from the conventional pulsar timing arrays. Pulsar timing arrays need a long-term observation of order $O(10)~{\rm years}$ to follow the wave form of nano-hertz GWs. Therefore, millisecond pulsars which have very stable periods through this time scale are necessary and only $5-10\%$ of millisecond pulsars satisfy this requirement. On the other hand, in the current case, we do not need such a long-term observation for two reasons. One is that we do not need to follow the wave form of the GWs and another is that the relative importance of linearly changing GWs and pulsar spin down in the timing residuals does not depend on the observation period (see Eqs. (\ref{eq:resh2}) and (\ref{eq:resp})). Although the pulsars still need to be stable enough for a period of a few years in order to determine $\dot{p}$ very precisely, the requirement on pulsar stability would be less strict. Thus, we expect that more millisecond pulsars could be used for this purpose compared with the conventional pulsar timing array, although it should be noted that only $\sim 30\%$ of millisecond pulsars have $\dot{p} < 10^{-20}$ and the increase would not be drastic.

\section{Summary and Discussion}

In this paper, we estimated gravitational waves from a hypothetical SMBH binary at the center of M87. This follows the scenario that the observed displacement between the AGN and the galactic center can be explained by the existence of another SMBH, or multiple IMBHs or small SMBHs. Because the period of the binary and the resulting GWs is much longer than observational time span, the amplitude changes linearly with time and we calculated the time derivatives of the GW amplitudes. We investigated their dependence on the orbital elements and the second BH mass taking the observational constraints into account.

There would be no chance to detect GWs from a SMBH binary which explains the observed offset of the AGN with either the conventional pulsar timing arrays or a new method proposed above. In the second scenario where multiple IMBHs and small SMBHs reside and one of them is forming a binary with the main SMBH, there is a small chance to detect the GWs with the new method. However, the detection of such GWs is not a direct evidence of the existence of multiple BHs and, hence, does not give a decisive explanation of the AGN offset. Nevertheless, linearly changing GWs would still be a useful tool to probe the center of M87.

If multiple BHs reside at the center of M87, their gravitational perturbations will affect not only the main SMBH but also the distribution and velocity field of stars \citep{Perets2007}. This could be another probe of the M87 center and will be studied in future.

\begin{ack}
KT is supported by Grand-in-Aid from the Ministry of Education, Culture, Sports, and Science and Technology (MEXT) of Japan, No. 24340048, No. 26610048 and No. 15H05896. The research of JS has been supported at IAP and at CEA Saclay by the ERC Project No.~267117 (DARK) hosted by Universit\'e Pierre et Marie Curie (UPMC) - Paris 6 and at JHU by NSF Grant No.~OIA-1124403.
\end{ack}

\end{document}